\documentstyle[preprint,eqsecnum,aps]{revtex}

\begin{document}
\draft \preprint{HEP/123-qed}
\title{Spectral Properties of the Su-Schrieffer-Heeger Model}
\author{Marco Zoli}
\address{Istituto Nazionale Fisica della Materia - Universit\'a di Camerino, \\
62032 Camerino, Italy. e-mail: zoli.marco@libero.it }

\date{\today}
\maketitle
\begin{abstract}
We present a study of the one dimensional  Su-Schrieffer-Heeger
model Hamiltonian by a diagrammatic perturbative method in the
weak electron-phonon coupling regime. Exact computation of both
the charge carrier effective mass and the electron spectral
function shows that electrons are good quasiparticles in the
adiabatic and antiadiabatic limits but novel features emerge in
the intermediate regime, where the phonons and the electrons
compare on the energy scale. Together with a sizeable mass
enhancement we observe, in the latter regime, a spread of the
spectral weight (among several transition peaks) associated with
an increased relevance of multiphonons contributions at larger
{\it e-ph} couplings. Accordingly electrons cease to be the good
quasiparticles and an onset of polaron formation is favoured.
\end{abstract}
\widetext
\section*{1.Introduction}

The coupling of the electrons to the quantum lattice vibrations is
a key issue in solid state physics. While in the weak coupling
limit electrons generally behave as good quasiparticles, in strong
coupling regimes electrons may drag the lattice deformation
induced by the scattering process. In the latter case, the charge
carrier is the unit comprising  electron and  dressing cloud of
phonons, namely the polaron \cite{landau}. A wide literature has
been produced on the subject during the last decades
\cite{toyozawa,eagles,emin,devreese,romero,mello} and particular
emphasis has been recently laid on the polaronic properties of
high $T_c$ superconductors \cite{mott}. Theoretical investigations
usually start from the Holstein molecular crystal model
\cite{holstein} which assumes a momentum independent coupling of
electrons to dispersive optical phonons  leading to polaron
formation if the energy gain associated with the lattice
deformation is larger than the kinetic energy due to the tight
binding hopping in the bare band \cite{io4}. As a remarkable
feature of the polaron physics \cite{rashba,raedt,kopida,tsiro},
the strength of the coupling drives the transition from a large
size polaron (at weak couplings) to a small size polaron (at
strong couplings). This transition \cite{gerlow} is accompanied by
a sizeable enhancement of the effective mass whose value  may
change considerably according to the degree of adiabaticity and
the peculiarities of the lattice structure \cite{io3,io2}. In some
systems however \cite{barisic,schulz,lu} the {\it e-ph}
interaction modifies the electron hopping matrix elements thus
leading to a momentum dependent coupling function. In this case an
appropriate theoretical framework is offered by the
Su-Schrieffer-Heeger(SSH) model Hamiltonian \cite{ssh} originally
proposed to explain the conducting properties of quasi one
dimensional polymers as polyacetylene \cite{ssh1} whose twofold
degenerate ground state sustains well known nonlinear local
excitations, the solitons. More generally, the SSH Hamiltonian
provides an alternative (to the Holstein model) tool to analyse
the physics of polaron formation as tuned by the strength of the
{\it e-ph} coupling in one dimensional systems. This physics has
been less deeply investigated than in the Holstein model. With the
present paper we address precisely this question attacking the SSH
Hamiltonian by a weak coupling perturbative method. Although this
approach is clearly not adequate to capture the multiphononic
nature of the polaronic quasiparticle, still it can provide useful
informations regarding the onset of polaron formation in some
portions of parameter space. Here we look first at the mass
renormalization exploring a wide range of values for the adiabatic
parameter and, successively, we compute the electronic spectral
function to detect whether and to which extent bare electrons
behave as good quasiparticles. The Section 2 outlines the SSH
model and contains the results of this study while some
conclusions are drawn in Section 3.

\section*{2.Model and Results}

In 1D, the real space SSH Hamiltonian reads

\begin{eqnarray}
H=\,& & \sum_{r}J_{r,r+1} \bigl(f^{\dag}_r f_{r+1} +
f^{\dag}_{r+1} f_{r} \bigr) +  \sum_{r}\Bigl({{p^2_r}\over {2M}} +
{K \over 2}(u_r - u_{r+1})^2
 \Bigr)\,
 \nonumber \\
& &J_{r,r+1}=\, - {1 \over 2}\bigl[ J + \alpha (u_r -
u_{r+1})\bigr] \label{1}
\end{eqnarray}

where $J$ is the nearest neighbors hopping integral for an
undistorted chain, $\alpha$ is the electron-phonon coupling, $u_r$
is the dimerization coordinate which specifies the displacement of
the $r-$ lattice site from the equilibrium position along the
molecular axis, $p_r$ is the momentum operator conjugate to $u_r$,
$M$ is the ion (ionic group) mass, $K$ is the effective spring
constant, $f^{\dag}_r$ and $f_{r}$ create and destroy electrons on
the $r-$ site. Let's expand the lattice displacement and its
conjugate momentum in terms of the phonon creation and
annihilation operators $b^{\dag}_{q}$ and $b_{q}$ and Fourier
transform the electron operators

\begin{eqnarray}
u_r=\,& &\sum_q {1 \over {\sqrt{2MN\omega_q}}} \bigl(b^{\dag}_{-q}
+ b_q \bigr)\exp(iqr)\,
 \nonumber \\
p_r=\,& & i \sum_q \sqrt{{M \omega_q} \over {2N}}
\bigl(b^{\dag}_{-q} - b_q \bigr)\exp(iqr)\,
 \nonumber \\
f_r=\,& & {1 \over {\sqrt{N}}}\sum_k \exp(ikr)f_k
 \label{2}
\end{eqnarray}

in order to obtain the SSH Hamiltonian in momentum space:

\begin{eqnarray}
& &H=\, H_0 + H_{int}\,
 \nonumber \\
& &H_0=\, \sum_{k}\varepsilon_k f^{\dag}_k f_k +  \sum_{q}
\omega_q b^{\dag}_q b_q \,
 \nonumber \\
& &H_{int}=\, \sum_{k,q} g(k+q,k) \bigl(b^{\dag}_{-q} + b_q \bigr)
f^{\dag}_{k+q} f_k \,
 \nonumber \\
& & \varepsilon_k=\, -J \cos (k) \,
 \nonumber \\
& & \omega^2_q=\, 4{{K \over M}}\sin^2 ({q \over 2}) \,
 \nonumber \\
& &g(k+q,k)=\, {{i \alpha} \over {\sqrt {2MN \omega_q}}}
\bigl(\sin (k+q) - \sin (k) \bigr) \label{3}
\end{eqnarray}

where $N$ is the total number of lattice sites. The phonon
dispersion relation is defined in the range $q \in [0,\pi]$.
Assuming a reduced Brillouin zone ($|q| \le \pi/2$) the spectrum
displays both an acoustic and an optical branch \cite{naka}. The
model contains three free parameters: the hopping integral $J$,
the zone boundary frequency $\omega_{\pi}=\,2\sqrt{{K/M}}$ which
coincides with the zone center  optical frequency in the reduced
zone scheme, the coupling constant $\alpha^2/4K$.

The full electron propagator in the Matsubara Green's functions
formalism is defined as:

\begin{equation}
G(k,\tau)=\,-\sum_{n=0}^\infty (-1)^n \int_0^\beta d\tau_1
...d\tau_n \Bigl < T_\tau f_k(\tau) H_{int}(\tau_1) \cdot \cdot
H_{int}(\tau_n) f^\dag_k(0) \Bigr >_0 \label{4}
\end{equation}

where $\beta$ is the inverse temperature, $T_\tau$ is the time
ordering operator, $<...>_0$ indicates that thermodynamic averages
are taken with respect to the unperturbed Hamiltonian and only
different connected diagrams contribute to any order $n$. I have
computed exactly the self-energy terms due to one phonon ($n=2$ in
eq.(4)) and two phonons ($n=4$ in eq.(4)) scattering processes
which determine the renormalized electron mass $m_{eff}$ through
the relations:

\begin{equation}
{{m_{eff}}\over {m_0}}= \,{{1 - {{\partial Re\Sigma_k(\epsilon)}/
{\partial \epsilon }}|_{k=0; \,\, \epsilon=-J}} \over {1 +
{{\partial Re\Sigma_k(\epsilon)}/ {\partial \varepsilon_k
}}}|_{k=0; \,\, \epsilon=-J}}
\label{5}
\end{equation}

where, $Re\Sigma_k(\epsilon)=\, Re\Sigma_k^{(1)}(\epsilon) +
Re\Sigma_k^{(2a)}(\epsilon) + Re\Sigma_k^{(2b)}(\epsilon) +
Re\Sigma_k^{(2c)}(\epsilon)$  is the frequency dependent real part
of the retarded self-energy. There are three contributions due to
different connected two-phonons diagrams \cite{mahan}. Their
effect is however confined to the intermediate regime in which
$\omega{_\pi}$ is comparable to the electronic energy $J$. Here
the two phonons diagrams enhance the effective mass by $\sim 15\%$
with respect to the one phonon result. Instead, in the fully
adiabatic and antiadiabatic regimes the two phonons contributions
(evaluated at the band bottom) are negligible. Hereafter, the
displayed results depend on the very one phonon self-energy
$\Sigma_k^{(1)}(i\epsilon_m)$ term ($\epsilon_m=(2m+1)\pi/\beta$
with $m$ integer number) whose finite temperatures analytic
expression is given by:

\begin{equation}
\Sigma_k^{(1)}(i\epsilon_m)=\,-\sum_q g^2(k,k-q) \Biggl[
{{n_B(\omega_q) + n_F(-\varepsilon_{k-q})}\over {i\epsilon_m -
\varepsilon_{k-q} - \omega_q}} + {{n_B(\omega_q) +
n_F(\epsilon_{k-q})}\over {i\epsilon_m - \varepsilon_{k-q} +
\omega_q}} \Biggr] \label{6}
\end{equation}

$n_B$ and $n_F$ are the Bose and Fermi occupation factors
respectively.
 We take a narrow band system setting $J=\,0.1eV$. Figure 1 shows a
sizeable mass enhancement in the intermediate adiabatic regime
with a pronounced spike at $\omega_{\pi} \sim \sqrt{2} J$ mainly
due to scattering by $|q|=\,\pi/2$-phonons. This feature is
related to the 1D electron and phonon dispersion relations and it
may be partly suppressed in higher dimensionality. However the
onset of a mass renormalization starting at $\omega_{\pi} \sim
J/2$ and, more evidently, at $\omega_{\pi} \sim J$ (together with
the increased relevance of multiphonons contributions) signals
that polaron formation is expected in this regime while no mass
enhancement is obtained in the adiabatic and antiadiabatic limits.
Let's look now at the spectral function defined by $A(k,
\,\epsilon)=\,-2Im G_{ret}(k, \,\epsilon)$ to get more insight
\cite{robin} into the suggestions proposed by the effective mass
behavior. In terms of the retarded self-energy, obtained by eq.(6)
through analytic continuation $\epsilon_m \to \epsilon + i
\delta$, $A(k, \,\epsilon)$ reads:

\begin{equation}
A(k, \,\epsilon)= \,2\pi \delta\bigl(\epsilon - \varepsilon_k -
Re\Sigma_k(\epsilon)\bigr) + {{(-) 2 Im \Sigma_k(\epsilon)} \over
{\Bigl(\epsilon - \varepsilon_k - Re\Sigma_k(\epsilon) \Bigr)^2 +
\Bigl[ Im \Sigma_k(\epsilon) \Bigr]^2}} \label{7}
\end{equation}

The first addendum in eq.(7) contributes when $Im
\Sigma_k(\epsilon)=\,0$. The band bottom ($k=\,0$) spectral
function has been computed in a number of representative cases
summing the self-energy term (eq.(6)) over 6000 $q-$points in the
Brillouin zone and searching the zeros of the $\delta$-function
argument (eq.(7)) and $Im \Sigma_{k=\,0}(\epsilon)$ over $10^5$
points of the $\epsilon-$ axis. The sum rule
$$\int_{-\infty}^{\infty}{{d\epsilon} \over {2\pi}} A(k=\,0,
\,\epsilon) =\,1$$ has to be numerically fulfilled in principle.
Here however we are approximating the total self-energy by the
one-phonon term (eq.(6)) which linearly depends on the free
parameter $\alpha^2/4K$. As the {\it e-ph} coupling grows
multiphonons terms become more relevant and our approximation
becomes less accurate. Accordingly, deviations from the sum rule
are expected as a measure of the loss of spectral weight
associated with higher order self-energy effects. In this regard,
the sum rule numerical analysis permits to define the range of
$\alpha^2/4K$ values within which the one phonon approximation is
reliable. For any choice of input parameters we are able to
estimate the intrinsic error of our physical model.

Figures 2(a) and 2(b) illustrate that electrons are good
quasiparticles in the fully antiadiabatic regime: for both values
of the {\it e-ph} coupling there is a well resolved peak due to
the $\delta-$function contribution and located at $\epsilon
=\,-98.8meV$ (Fig.2(a)) and $\epsilon =\,-90.4meV$ (Fig.2(b)),
respectively.

The sum rule is well satisfied in Figure 2(a) while a slight loss
of spectral weight ($8\%$) occurs in Figure 2(b) (where the
dimensionless effective coupling is $\alpha^2/(4KJ)=\,0.1$) ,
being $\int_{-\infty}^{\infty}{d\epsilon} A(k=\,0,
\,\epsilon)/{2\pi} =\,0.92$. Infact, by increasing the {\it e-ph}
coupling fourth order diagrams acquire relevance and should
contribute to the fulfillement of the sum rule. Figures 3 deal
with the intermediate regime $\omega_{\pi}=\,J$. Electrons are
still good excitations in the extremely weak coupling case
(Fig.3(a)) with one well defined transition at $\epsilon
=\,-100.9meV$ but, at larger couplings (Fig.3(b)), the sum rule is
far from being satisfied and a $40\%$ loss of spectral weight is
observed together with a strong reduction of the main peak height.
Here we see the failure of the electronic quasiparticle picture
and the possible onset of a polaronic state. Let's come to the
adiabatic regime represented in Figures 4. In Figure 4(a) the sum
rule is satisfied and the spectral weight spreads mainly in a few
peaks around the highest one located at the energy $\epsilon
=\,-100meV$ with a significant tail up to energy levels of order
$\epsilon \sim \,-80meV$. At larger couplings (Fig.4(b)) the tail
is appreciable up to $\epsilon \sim \,-50meV$ and the loss of
spectral weight is $\sim 20\%$. A well defined transition is
however still present at $\epsilon =\,-108.2meV$. The temperature
induced effects on the spectral function of the extremely weak
coupling system are shown in Figures 5(a) for the $T=100K$ case
and 5(b) for the $T=300K$ case, respectively. Comparison with the
ground state case (Figure 4(a)) emphasizes the progressive
broadening of the main peak in a quasi continuum of states which,
above room temperature, tend to have similar probabilities.

\section*{3. Final Remarks}

Electrons are expected to be good quasiparticle and natural
excitations of a system in which the electron-phonon interaction
is weak. We have studied the Su-Schrieffer-Heeger Hamiltonian by a
low order perturbative method assuming a coupling energy
$\alpha^2/4K$ always smaller than the bare hopping integral $J$
which sets the scale of the electron energies. The effective mass
of the charge carriers varies with the degree of adiabaticity of
the system: while no renormalization is observed in the adiabatic
and antiadiabatic limits, a sizeable enhancement takes place when
the phonons compare in energy with the electrons, namely in the
intermediate regime. Although the pronounced mass peak may be
partly reduced in higher dimensionality exact computations of
fourth order diagrams clearly show that multiphonons effects
become relevant in the intermediate regime thus signalling that a
polaronic behavior may here occur. Analysis of the electronic
spectral function corroborates these findings. More precisely than
the effective mass, the electron spectral function, directly
measures to which extent electrons are the good excitations of the
system in any regime tuned by the adiabaticity parameter. Only one
phonon scattering processes have been considered in the presented
results. The spectral function sum rule has been assumed as a
benchmark for the reliability of the one phonon approximation and
the computed loss of spectral weight have been associated with an
increased relevance of multiphonons contributions  as a function
of the strength of the {\it e-ph} coupling.

While, in the antiadiabatic regime, the system presents well
defined electronic transitions again, in the intermediate regime,
electrons cease to be good quasiparticles as soon as the {\it
e-ph} coupling grows, although well within a weak coupling scheme.
This is an independent confirmation of the onset of polaronic
features suggested by the mass calculation. Also in the moderately
adiabatic regime we find that the spectral weight spreads over a
region of energy levels but a defined transition is present in the
ground state of the system. The polaronic properties of the model
will be next studied in the strong coupling regime thus permitting
a comparison with the Holstein molecular crystal model.
Dimensionality effects will be also investigated.

\begin{figure}
\vspace*{8truecm} \caption{Renormalized masses (in units of bare
band electron mass) versus the adiabaticity parameter.
$m_{eff}^{(1)}$ is due to the one phonon self-energy correction.}
\end{figure}

\begin{figure}
\vspace*{8truecm} \caption{Electron spectral function in
antiadiabatic regime and (a) extremely weak {\it e-ph} coupling;
(b) moderately weak {\it e-ph} coupling. }
\end{figure}

\begin{figure}
\vspace*{8truecm} \caption{Electron spectral function in
intermediate regime and (a) extremely weak {\it e-ph} coupling;
(b) moderately weak {\it e-ph} coupling.}
\end{figure}

\begin{figure}
\vspace*{8truecm} \caption{Electron spectral function in adiabatic
regime and (a) extremely weak {\it e-ph} coupling; (b) moderately
weak {\it e-ph} coupling. }
\end{figure}

\begin{figure}
\vspace*{8truecm} \caption{Finite temperatures electron spectral
function in adiabatic regime and  extremely weak {\it e-ph}
coupling; (a) $T=\,100K$, (b) $T=\,300K$.}
\end{figure}

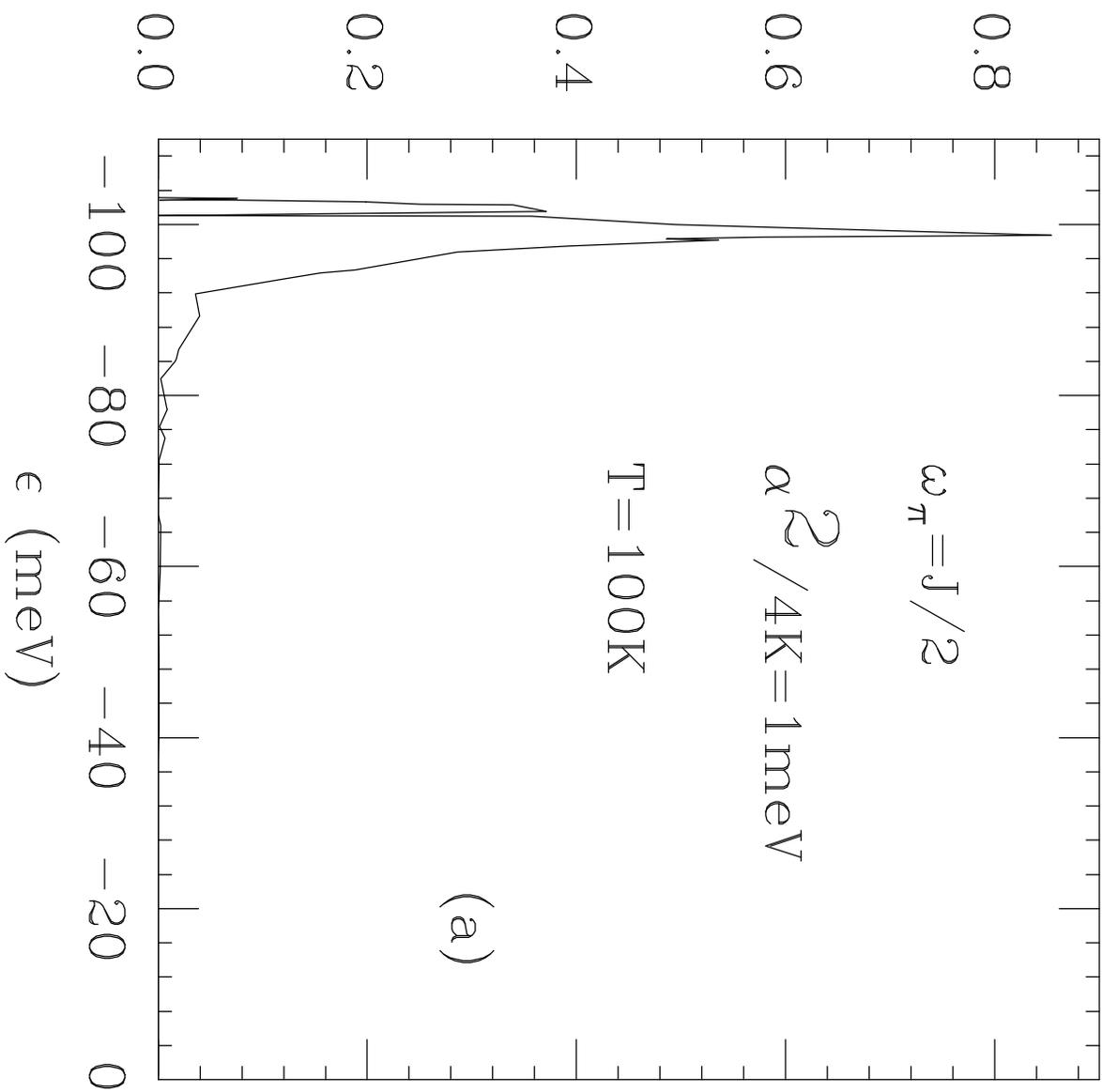
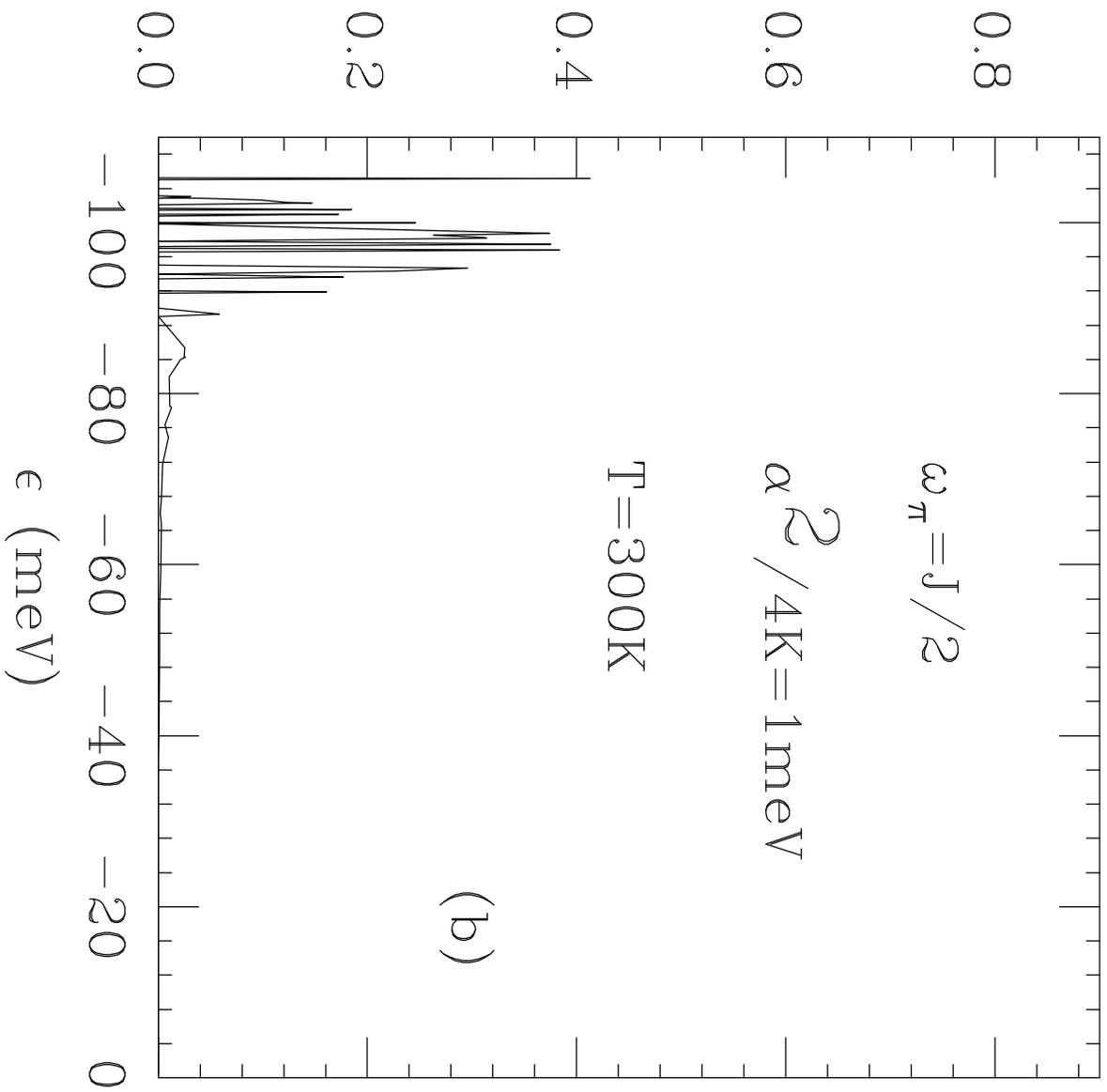
\end{document}